\documentclass[a4paper,12pt,twocolumn]{article}

\title{ Faint young Sun paradox remains}
\author{Colin Goldblatt$^{1,2}$ and Kevin J. Zahnle$^{1}$}
\date{}

\begin{document}

\maketitle

$^{1}$ Space Science and Astrobiology Division, NASA Ames Research Center,
MS 245-3, Moffett Field, California 94035, USA. 

$^{2}$ Present address: Astronomy Department and Virtual Planetary
Laboratory, University of Washington, Box 351580, Seattle, Washington
98195, USA. cgoldbla@uw.edu

\begin{center}
\textbf{Brief Communication Arising}\newline
Arising from: M. T. Rosing, D. K. Bird, N. H. Sleep \& C. J. Bjerrum, Nature 464, 744--747 (2010).                                                                        \end{center}

The Sun was fainter when the Earth was young, but the climate was
generally at least as warm as today; this is known as the `faint young
Sun paradox'. Rosing et al. [1] claim that the paradox can be resolved by
making the early Earth's clouds and surface less reflective. We show
that, even with the strongest plausible assumptions, reducing cloud
and surface albedos falls short by a factor of two of resolving the
paradox. A temperate Archean climate cannot be reconciled with
the low level of CO$_2$ suggested by Rosing et al. [1]; a stronger greenhouse
effect is needed.

During the Archean eon, the Earth received 76\% to 83\% of the
energy from the Sun that it does today. If the Earth's greenhouse effect
and albedo were the same as now, the Earth would have been in
continual deep freeze until one billion years ago, with glaciers reaching
the Equator. However, Archean glacial sediments are rare and
geological evidence indicates that the Archean was typically warmer
than today (we are in a glacial period now). With the amount of
energy reaching the Earth given by $F = \frac{1}{4}(1-\alpha) = 239$\,Wm$^2$
(using the present-day solar constant $S = 1368$\,Wm$^2$ and albedo
$\alpha = 0.3$), the radiative deficit in the Archean would have been $(1-
0.79)F<50$\,Wm$^2$. Resolution of the `faint young Sun paradox'
requires a positive radiative forcing---from reducing the albedo or
increasing the greenhouse effect---of more than 50\,Wm$^2$.

Clouds have two competing radiative effects: they reflect sunlight
but they also add to the greenhouse effect if they are colder than the
surface. Reflection dominates in low clouds, and the greenhouse effect
dominates in high clouds. Therefore the absolute upper bound on
warming by decreasing cloud reflectivity would be found by removing
low clouds entirely. This gives a forcing of 25\,Wm$^2$, half of what is
needed to resolve the `faint young Sun paradox' (our cloud model is
described in the Methods). Any reduction to high clouds would cause
a cooling.

Rosing et al. [1] justify less-reflective clouds with the incorrect statements
that most cloud condensation nuclei (CCN) are from biogenic dimethyl
sulphide (DMS), and that DMS is solely produced by eukaryotes. DMS
is also produced microbially [2]. Products of DMS contribute only 3\% of
Northern Hemisphere CCN and 10\% of Southern Hemisphere CCN
today [3]. Other biological [4] and non-biological sources, especially sea salt,
provide CCN. If CCN production were to depend only on eukaryotic
DMS emissions1, we would expect to see significant cooling when
eukaryotes evolved, but no such cooling is evident.

Nevertheless, we can assume no biological CCN supply and quantify
the resulting forcing. Over the modern ocean the effective radius $r_e$ of
cloud drops rarely exceeds 15\,$\mu$m [5] even in remote and unproductive
regions (the $r_e$ of 17\,$\mu$m to 30\,$\mu$m used by Rosing et al. [1] is too
high). For an upper bound, we increase low cloud droplet size by 50\%
from our standard case, from 11\,$\mu$m to 16.5\,$\mu$m. With no change in cloud
thickness, the forcing is 7\,Wm$^2$. Clouds with larger drops may rain out
faster. Parameterizations of enhanced rain-out vary from proportional
to $(r_{e,0}/r_e)^1$ to proportional to $(r_{e,0}/r_e)^5.37$ [6,7]; the corresponding
extra forcing would be 4--15\,Wm$^2$ (remote sensing data for marine
stratus suggest that the low end of this range is more appropriate [8]). The
sum is 11\,Wm$^2$ to 22\,Wm$^2$, with the low end being most likely.

The authoritative estimate of the global energy budget [9] gives global
mean and ocean albedos of 0.125 and 0.090 respectively. The largest
realistic surface darkening is from the present mean to an all-ocean
world, which gives a radiative forcing of 5\,Wm$^2$.

Increasing the CO$_2$ mixing ratio to 1000 parts per million by
volume (p.p.m.v.; the upper bound according to Rosing et al. [1]) gives
a forcing of 6\,Wm$^2$. Rosing et al. [1] rely on 1000 p.p.m.v. CH$_4$ for
much of their warming, ignoring relevant atmospheric chemistry. As
the partial pressure of CH$_4$ (pCH$_4$ ) approaches that of CO$_2$ (pCO$_2$ ),
hydrocarbon haze forms in the stratosphere, the cooling effect of
which outweighs the greenhouse effect of CO$_2$ and CH$_4$ [10,11]. Numerical models [12] predict haze production when pCH$_4$ /
pCO$_2$ 50.1 and haze production has been seen in laboratory experiments13
where $p\mathrm{CH}_4 / p\mathrm{CO}_2 = 0.3$. With 1000 p.p.m.v. CO$_2$, the maximum
CH$_4$ concentration that can give warming is 300 p.p.m.v., which
would contribute 7\,Wm$^2$ of additional forcing.

Changes to clouds could in theory considerably reduce the amount
of greenhouse gases required, because gaseous absorption depends on
the logarithm of gas abundance. But even with the highly unlikely
assumption of no biological CCN supply, cloud changes can provide
only one-quarter to one-half of the required radiative forcing. Any
changes to clouds would require strong justification, which Rosing et
al. [1] do not provide. A strong greenhouse effect is required in the Archean. The alternative is an extremely cold climate with continual mid- to low-latitude glaciation, for which there is no evidence

\textbf{Methods} We calculate the radiative forcing (change in net flux at the tropopause) on a
single global annual mean atmospheric profile, with three layers of clouds that
overlap randomly [14]. The radiative fluxes on eight sub-columns corresponding to
each cloud combination are calculated with the RRTM model [15]. For our standard
case, cloud water paths are [$W_\mathrm{high}$,$W_\mathrm{mid}$,$W_\mathrm{low}$] = [20, 25, 40]\,gm$^2$, fractions are
[$f_\mathrm{high}$,$f_\mathrm{mid}$,$f_\mathrm{low}$] = [0.25, 0.25, 0.40] and the surface albedo is 0.125. Standard lowand
mid-level clouds are liquid with $r_e = 11$\,$\mu$m and high clouds are ice with
generalized effective size of $D_{ge} = 75$\,$\mu$m. For radiative forcings described in the
text, the low cloud water path is varied but all other parameters are unchanged.

\vspace{0.4cm}

\begin{small}
1. Rosing, M. T., Bird, D. K., Sleep,N.H. \& Bjerrum, C. J. No climate paradox under the
faint early Sun. \textit{Nature} 464, 744--747 (2010).

2. Lin, Y. S., Heuer, V. B., Ferdelman, T. G. \& Hinrichs, K.-U. Microbial conversion of
inorganic carbon to dimethyl sulfide in anoxic lake sediment (Plussee, Germany).
\textit{Biogeosciences} 7, 2433--2444 (2010).

3. Woodhouse, M. T. et al. Low sensitivity of cloud condensation nuclei to changes in
the sea-air flux of dimethyl-sulphide. \textit{Atmos. Chem. Phys}. 10, 7545--7559 (2010).

4. Leck, C. \& Bigg, E. K. Source and evolution of the marine aerosol--a new
perspective. \textit{Geophys. Res. Lett.} 32, L19803, doi:10.1029/2005GL023651
(2005).

5. Breon, F. Tanre, D. \& Generoso, S. Aerosol effect on cloud droplet size, monitored
from satellite. \textit{Science} 295, 834--838 (2002).

6. Kump, L. R. \& Pollard, D. Amplification of Cretaceous warmth by biological cloud
feedbacks. \textit{Science} 320, 195 (2008).

7. Penner, J. E. et al. Model intercomparison of indirect aerosol effects. \textit{Atmos. Chem.
Phys.} 6, 3391--3405 (2006).

8. Kaufman, Y. J., Koren, I.,Remer, L. A., Rosenfeld, D. \& Rudich, Y. The effect of smoke,
dust, and pollution aerosol on shallow cloud development over the Atlantic Ocean.
\textit{Proc. Natl Acad. Sci. USA} 102, 11207--11212 (2005).

9. Trenberth, K. E., Fasullo, J. T. \& Kiehl, J. T. Earth's global energy budget. \textit{Bull. Am.
Meteorol. Soc.} 90, 311--323 (2009).

10. McKay, C. P., Lorenz, R. D. \& Lunine, J. I. Analytic solutions for the anti-greenhouse
effect: Titan and the early Earth. \textit{Icarus} 91, 93--100 (1991).

11. Haqq-Misra, J. D., Domagal-Goldman,S.D., Kasting,P. \& Kasting, J.F. A revised,hazy
methane greenhouse for the Archean Earth. \textit{Astrobiology} 8, 1127--1137 (2008).

12. Domagal-Goldman, S. D., Kasting, J. F., Johnston, D. T. \& Farquhar, J. Organic haze,
glaciations and multiple sulfur isotopes in the Mid-Archean Era. \textit{Earth Planet. Sci.
Lett.} 269, 29--40 (2008).

13. Trainer, M. G. et al. Organic haze on Titan and the early Earth. \textit{Proc. Natl Acad. Sci.
USA} 103, 18035--18042 (2006).

14. Goldblatt, C. \& Zahnle, K. J. Clouds and the faint young Sun paradox. \textit{Clim. Past} 7,
203--220 (2011).

15. Clough, S. A. et al. Atmospheric radiative transfermodeling: a summary of the AER
codes. \textit{J. Quant. Spectrosc. Radiat. Transf.} 91, 233--244 (2005).
\end{small}

\vspace{0.4cm}

\textbf{Author Contributions:} C.G. and K.J.Z. discussed the article to which this note responds. C.G. performed all model runs and quantitative analysis. C.G. and K.J.Z. both contributed qualitative analysis and both contributed to writing the paper.
  
\textbf{Competing financial interests:} declared none.

\textbf{doi:}10.1038/nature09961

\end{document}